\documentclass[a4paper, 11pt, oneside]{article}

\usepackage{jheppub}

\newcommand\fverb{\setbox\fverbbox=\hbox\bgroup\verb}
\newcommand\fverbdo{\egroup\medskip\noindent%
            \fbox{\unhbox\fverbbox}\ }
\newcommand\fverbit{\egroup\item[\fbox{\unhbox\fverbbox}]}
\newbox\fverbbox

\title{$\sin{\theta}_{13}$ and neutrino mass matrix with an approximate flavor symmetry}
\author[1]{Riazuddin}
\affiliation[1]{National Centre for Physics,\\
Quaid-i-Azam University Campus, Islamabad 45320, Pakistan }
\emailAdd{riazuddin@ncp.edu.pk}
\date{\today}

\abstract{For a neutrino mass matrix whose texture has an approximate flavor
symmetry and where one has near degenerate neutrino mass, it is shown
that the tribimaximal values for atmospheric angle
$\sin^{2}{\theta_{23}}=\frac{1}{2}$ and solar angle
$\sin^{2}{\theta_{12}}=\frac{1}{3}$ can be maintained even when the
reactor angle $\theta_{13}\neq0$. The non zero $\sin\theta_{13}$ implies approximate
$\nu_{\mu}\rightarrow-\nu_{\tau}$ symmetry instead of
$\nu_{\mu}\rightarrow\nu_{\tau}$ symmetry.  }

\keywords{Neutrino-Physics, Flavor Symmetry, Nonzero reactor angle}

\begin{document}
\maketitle

\section{Introduction}\label{intro}
There is compelling evidence that neutrinos change flavor, have non
zero masses and that neutrino mass eigenstates are different from
weak eigenstates. As such they undergo oscillations. The flavor and
mass eigenstates are related by the so called
Pontecorvo-Maki-Nakagawa-Sakata (PMNS) mixing matrix \cite{1},
\begin{eqnarray}
\left(
\begin{tabular}{c}
$\nu _{e}$ \\
$\nu _{\mu }$ \\
$\nu _{\tau }$%
\end{tabular}%
\right) =U\left(
\begin{tabular}{c}
$\nu _{1}$ \\
$\nu _{2}$ \\
$\nu _{3}$%
\end{tabular}%
\right),   \label{1}
\end{eqnarray}
where the matrix $U$ has been parameterized by the Particle Data
Group (PDG) as \cite{2}

\begin{equation}
\left(
\begin{tabular}{ccc}
$c_{12}c_{13}$ & $s_{12}c_{13}$ & $s_{13}e^{-i\delta }$ \\
$-s_{12}c_{23}-c_{12}s_{23}s_{13}e^{i\delta }$ & $%
c_{12}c_{23}-s_{12}s_{23}s_{13}e^{i\delta }$ & $s_{23}c_{13}$ \\
$s_{12}s_{23}-c_{12}c_{23}s_{13}e^{i\delta }$ & $%
-c_{12}s_{23}-s_{12}c_{23}s_{13}e^{i\delta }$ & $c_{23}c_{13}$%
\end{tabular}%
\right)P=U_{PMNS} P .   \label{2}
\end{equation}%
Here $c_{ij}=\cos{\theta_{ij}}$, $s_{ij}=\sin{\theta_{ij}}$ and $P$
is a diagonal matrix which contains (Majorana) CP violating phases
in addition to the (Dirac) CP violating phase $\delta$.
$\theta_{12}$, $\theta_{23}$ and $\theta_{13}$ are respectively
known as solar, atmospheric and reactor angles.

Current global fits allow the following ranges for the mass squared
differences and mixing angles \cite{2}:
\begin{eqnarray}
7.05\times10^{-5}eV^{2}&\leq\Delta
m_{12}^{2}\leq&8.34\times10^{-5}eV^{2}, \nonumber \\
0.25&\leq \sin^{2}{\theta_{12}}\leq&0.37, \nonumber \\
2.70\times10^{-3}eV^{2}&\leq\Delta
m_{31}^{2}\leq&2.75\times10^{-3}eV^{2}, \nonumber \\
0.36&\leq \sin^{2}{\theta_{23}}\leq&0.67,
 \label{3}
\end{eqnarray}
with the following best fit (BF) values
\begin{equation}
\Delta m_{12}^{2}=7.65\times10^{-5}eV^{2}, \ \ \
\sin^{2}{\theta_{12}}=0.304, \ \ \ \Delta
m_{31}^{2}=2.40\times10^{-3}eV^{2}, \ \ \ \sin^{2}{\theta_{23}}=0.5.
 \label{4}
\end{equation}
Recent results from T2K collaboration \cite{3} and MINOS indicate a
relatively large $\theta_{13}$ and when combined with the global fit
gives \cite{4}
\begin{equation}
\sin^{2}{\theta_{13}}=0.025\pm0.007.
 \label{5}
\end{equation}
There is further evidence for nonzero reactor also $\theta_{13}$ from DAYA BAY\cite{5} and RENO\cite{6} collaborations which respectively give
\begin{equation}
\theta_{13}=\left(8.83^{+0.81}_{-0.88}\right)^o \qquad \theta_{13}=\left(9.36^{+0.88}_{-0.96}\right)^o\notag
\end{equation}
It is interesting on its own right to consider
non-zero value for $\sin^{2}{\theta_{13}}$ in the above range. In fact it has been shown by the author\cite{7}
that nonzero value of $\sin^2{\theta_{13}}$ has important implications for the leptogenesis asymmetry parameter;
its contribution to this parameter may even dominate.

Before we proceed further it is instructive to summarize the theoretical framework needed.
The effective Majorana neutrino mass matrix $M_\nu$ constructed directly or in seesaw mechanism, can be symbolically written as \cite{8}
\begin{equation}
L=-\bar{L_\ell}M_\ell e_R-\bar{L_\ell}M_D N_R-N_R^T M_R N_R
\end{equation}
where $L_{\ell}=(e_L, v_L)$ are lepton doublets, $e_R$ charged
lepton $SU_L(2)$ singlets with non-vanishing hypercharge, $N_R$ are
$SU_L(2)\times U(1)$ singlets. It is convenient to have a basis in
which $M_\ell$ and $M_R$ are simultaneously diagonal
\begin{equation}
M_{\ell}\rightarrow \hat{M_\ell}=U_L^{\dag} M_\ell U_R,\,\,\,\, M_R\rightarrow \hat{M_R}=V^T M_R V
\end{equation}
Correspondingly $L_{\ell}\rightarrow U_{L} L_{\ell},
e_R\rightarrow U_R e_R$.

We can select a basis in which $U_L$ is diagonal i.e.
$M_\ell$=diag$(m_e, m_\mu, m_\tau)$. One may remark that the so called $2-3(\mu-\tau)$ symmetry can
not be simultaneously valid for left-handed charged leptons and
left-handed neutrinos. In the above basis it is obvious since
$m_\mu\neq m_\tau$ but in fact it is independent of what basis one
chooses \cite{9}. Thus $2-3(\mu-\tau)$ symmetry can only be regarded
as an effective (approximate) symmetry in the neutrino sector and
was inspired by maximal atmospheric angle and $\theta_r=0$. If
$\theta_r\neq 0$, it has to be violated.

Since the oscillation data are only sensitive to mass squared
differences, they allow for three possible arrangements of different
mass levels \cite{8}. It is customary to order the mass eigenstates
such that $m_1^2<m_2^2$. We have two squared mass differences:

$\Delta m^{2}_{21}=m^{2}_{2}-m^{2}_{1}=\delta m^{2}$, $\Delta
m^{2}=m_{3}^{2}-\frac{m^{2}_{2}+m^{2}_{1}}{2}$, where $\Delta
m^{2}>0$ for normal hierarchy ($m_{1}$ $\lesssim m_{2}$ $\leq
m_{3}$) and $<0$ for inverted hierarchy $(m_{1}\simeq
m_{2}\gg m_{3})$. For degenerate case $(m_{1}\simeq m_{2}\simeq
m_{3})$, one can write neutrino mass matrix as
\begin{eqnarray}
 M_{\nu}=m_{0}I+\delta M_{\nu} \label{7}
\end{eqnarray}
where $\delta M_{\nu}\ll m_{0}$ or
\begin{equation}
M_\nu=m_o\left(
        \begin{array}{ccc}
          1  &  0  &  0 \\
          0  &  0  &  1 \\
          0  &  1  &  0 \\
        \end{array}
      \right)+\delta M_\nu \label{11}
\end{equation}
in case of opposite CP-parity of $\nu_2$ and $\nu_3$ ($m_2$ and $m_3$ are of opposite sign).

We have three mixing angles $\theta_{12}=\theta_s$ (solar),
$\theta_{23}=\theta_a$ (atmospheric) and $\theta_{13}=\theta_r$
(reactor). Some new ingredients are needed to describe correctly the
three mixing angles. However it is well known that the best fit
values given in Eq. (\ref{4}) are consistent with the so called
tribimaximal (TB) mixing \cite{10} corresponding to

\begin{equation}
\sin^{2}{\theta_{12}}=\frac{1}{3}, \ \ \
\sin^{2}{\theta_{23}}=\frac{1}{2}, \ \ \ \sin^{2}\theta_{13}=0.\label{12}
\end{equation}

For our discussion it is
convenient to state various symmetries and/or conditions on
$M_{\nu}$ which lead to $\theta_{13}=0$ and TB mixing. In an
obvious notation if one has 2-3 symmetry i.e.
$(M_{\nu})_{22}=(M_{\nu})_{33}$ and $(M_{\nu})_{12}=(M_{\nu})_{13}$,
then $\theta_{13}=0$ and $\sin^{2}\theta_{23}=1/2$. If further
$(M_{\nu})_{11}+(M_{\nu})_{12}=(M_{\nu})_{22}+(M_{\nu})_{23}$, then
we have TB mixing given in Eq. (1.12).

Recently, a possibility has been discussed \cite{11}(named TBR) which
allows the extension of TB mixing, so as to have a non-zero value of
$\theta_{13}$, preserving at the same time the predictions for the
TB solar angle [$\sin^{2}{\theta_{12}}=\frac{1}{3}$] and the maximal
atmospheric angle [$\sin^{2}{\theta_{23}}=\frac{1}{2}$].

To implement TBR, it is generally assumed that
\begin{equation}
M_\nu^{TBR}=M_\nu^{TB}+\delta M_\nu\label{12}
\end{equation}
$M_\nu^{TB}$ satisfies the conditions mentioned above. The various
recent attempts in this aspect differ in the treatment of $\delta
M_\nu$. In general $\delta M_\nu$ contains six parameters. In
\cite{12}, the conditions on these parameters are put so as to give
$\theta_r\neq 0$ but $|\theta_{23}-45|\ll1$, fixing $\theta_s$ at
the TB value. However these conditions are not unique. A simple form
for $\delta M_\nu$ for normal hierarchy is also invented\cite{12}. In
\cite{13,14}, $\delta M_\nu$ is realized in specific flavor symmetry
models based on $S_4$ and $A_4$ symmetries respectively.

In this paper we attempt to realize TBR in degenerate mass spectrum
given in Eq. (\ref{7}) which as we shall see has some attractive
features.

\section{Approximate flavor symmetry and diagonalization of neutrino mass matrix}

For degenerate case a particularly attractive Majorana neutrino mass
matrix, which is a multiple of unit matrix
supplemented by three far smaller off-diagonal entries, is given by
\cite{15,16}
\begin{equation}
M_{\nu }=m_{0}\left(
\begin{tabular}{ccc}
$1$ & $\epsilon_{12}$ & $\epsilon_{13}$ \\
$\epsilon_{12}$ & $1$ & $\epsilon_{23}$ \\
$\epsilon_{13}$ & $\epsilon_{23}$ & $1$
\end{tabular}
\right)=m_{0}I+\delta M_{\nu}, \label{8}
\end{equation}
with $\epsilon_{ij}\ll1$. This implies that in the limit of off-diagonals going to zero, $M_\nu$ is flavor blind or preserves flavor
symmetry. When they are switched on, but being at least an order of magnitude smaller than diagonal elements, they violate it as small perturbations

A particularly attractive realization of $\delta \mathit{M_\nu}$ given in Eq.
(\ref{8})is provided by Zee model \cite{17} where diagonal elements
are vanishing and off-diagonal elements arise from radiative
corrections\cite{18}.

Another realization of $\delta M_\nu$ in Eq.(\ref{8}) is provided by a simple extension of the standard electroweak group to \cite{16}
\begin{equation}
 G\equiv S U_L(2)\times U_\ell (1) \times U_\mu(1)\times U_\tau (1)\notag
\end{equation}

In addition to usual fermions there are three right-handed
$SU_L(2)$ singlet neutrinos which carry appropriate $U_i(1)$ quantum
numbers. Further in addition to $SU_L(2)$ Higgs doublets, there are
three Higgs $SU_L(2)$ singlets $\Sigma^i$ with appropriate $U_i(1)$
quantum numbers. The fermions and Higgs bosons are assigned to the following representations of the group $G$:

\begin{equation}
L^{(1)}= L_e=\left[\begin{array}{c}
                \nu_e\\
                e
               \end{array}\right]_L
:(2,-1,0,0):\phi^{(1)}\notag
\end{equation}
\begin{equation}
 e_R:(1,-2,0,0), \qquad N_e:(1,-1,1,0);\qquad \Sigma^{(1)}:(1,1,-1,0)\notag
\end{equation}

\begin{equation}
L^{(2)}= L_\mu=\left[\begin{array}{c}
                \nu_\mu\\
                \mu
               \end{array}\right]_L
:(2,0,-1,0):\phi^{(2)}\notag
\end{equation}
\begin{equation}
 \mu_R:(1,0,-2,0), \qquad N_\mu:(1,1,0,-1);\qquad \Sigma^{(2)}:(1,-1,0,1)\notag
\end{equation}

\begin{equation}
L^{(3)}= L_\tau=\left[\begin{array}{c}
                \mu_\tau\\
                \tau
               \end{array}\right]_L
:(2,0,0,-1):\phi^{(3)}\notag
\end{equation}
\begin{equation}
 \tau_R:(1,0,0,-2), \qquad N_\tau:(1,0,-1,1);\qquad \Sigma^{(3)}:(1,0,1,-1)
\end{equation}
The Yukawa couplings of neutrinos with Higgs are given by
\begin{align}
 L_Y=&\sum_{i=e,\mu,\tau}g_i(\bar{L}^{(i)}\tilde{\phi}^{(i)}e_R^{(i)}+\text{H.C.})\notag\\
 &+\sum_{i=d,s,b}G_i(\bar{q}_L^{(i)}\tilde{\phi}^{(i)}q_R^{(i)}+\text{H.C.})\notag\\
 &+\sum_{i=u,c,t}G_i(\bar{q}_L^{(i)}\phi^{(i)}q_R^{(i)}+\text{H.C.})\notag\\
 &+(h_{11}\bar{L}_e N_e\phi^{(2)}+h_{23}\bar{L}_\mu N_\tau\phi^{(3)}+h_{32}\bar{L}_\tau N_\mu\phi^{(1)}+\text{H.C.})\notag\\
 &+\left[f_{12}(N_e^T c N_\mu+N_\mu^T c N_e)\tilde{\Sigma}^{(3)}\right.\notag\\
 &\left.+f_{13}(N_e^T C N_\tau+ N_\tau^T C N_e)\tilde{\Sigma}^{(2)}\right.\notag\\
 &\left.+f_{23}(N_\mu^T C N_\tau+ N_\tau^T C N_\mu)\tilde{\Sigma}^{(1)}+\text{H.C.}\right]
\end{align}
where
\begin{equation}
 \phi=\left( \begin{tabular}{c}
            $\phi^o$ \\
             $-\phi$
            \end{tabular}\right),\qquad
 \tilde{\phi}=-i \tau_2\phi^*=\left(\begin{tabular}{c}
                                  $\phi^\dag$\\
                                   $\phi^o$
                                 \end{tabular}\right)\notag
\end{equation}
The symmetry is spontaneously broken by giving vacuum expectation values to Higgs bosons $\phi^{(i)},\Sigma^{(i)}$:
\begin{equation}
 \langle\phi^{(i)}\rangle=\frac{v_i}{\sqrt{2}},\qquad
 \langle\Sigma^{(i)}\rangle=\frac{\Lambda_i}{\sqrt{2}}\notag
\end{equation}
where $\Lambda_i\gg v_i$ so that extra gauge bosons which break the $e-\mu-\tau$ universality become
super heavy and so do the right handed neutrinos. For simplicity we shall take, $v_1=v_2=v_3$ and $\Lambda_1=\Lambda_2=\Lambda_3$
(any differences can be absorbed in the corresponding Yukawa coupling constants with the Higgs bosons). Then the charged
lepton and Dirac neutrino mass matrices are
\begin{equation}
 M_\ell=\frac{v}{\sqrt{2}}\left(
 \begin{tabular}{ccc}
  $g_1$ & $0$ & $0$ \\
  $0$ & $g_2$ & $0$ \\
  $0$ & $0$ & $g_3$
 \end{tabular}\right)
\end{equation}
\begin{equation}
 m_D=\frac{v}{\sqrt{2}}\left(\begin{tabular}{ccc}
                              $h_{11}$ & $0$ & $0$\\
                              $0$ & $0$ & $h_{23}$\\
                              $0$ & $h_{32}$ & $0$
                             \end{tabular}\right)
\end{equation}
while $M_R$ is
\begin{equation}
 M_R=\frac{1}{\sqrt{2}}\left(\begin{tabular}{ccc}
                              $0$ & $f_{12}\Lambda_1$ & $f_{13}\Lambda_1$\\
                              $f_{12}\Lambda_1$ & $0$ & $f_{23}\Lambda_1$\\
                              $f_{13}\Lambda_1$ & $f_{23}\Lambda_1$ & $0$
                             \end{tabular}\right)
\end{equation}
Then the effective Majorana mass matrix for the light neutrinos is
\begin{equation}
 \delta M_\nu=\hat{m}_D \hat{M}_R^{-1} \hat{m}_D^T\label{131}
\end{equation}
where $\hat{m}_D$ is the Dirac matrix in $(\bar{N}_1 \quad \bar{N}_2 \quad \bar{N}_3)\left
          (\begin{tabular}{c}
                      $\nu_e$\\
                      $\nu_\mu$\\
                      $\nu_\tau$
                     \end{tabular}\right)$
basis giving
\begin{eqnarray}
 \delta M_\nu=\frac{v^2}{2 M_1}\left(\begin{tabular}{ccc}
                            $-h_{11}^2$ & $h_{11}h_{23}$ & $h_{11}h_{32}$\\
                            $h_{23}h_{11}$ & $-h_{23}^2$ &$h_{23}h_{32}$\\
                            $h_{32}h_{11}$ & $h_{23}h_{32}$ & $-h_{32}^2$
                           \end{tabular}\right)\label{121}
\end{eqnarray}
Here $M_1=f_1\displaystyle{\frac{1}{\sqrt{2}}}$ taking all Yukawa couplings in $M_R$ equal.

In order to generate $M_\nu^0$, we introduce a right handed neutrino $N$ and a corresponding Higgs boson $\Sigma$, both of
which are $S U_L(2)$ and $[U(1)]^3$ singlets, with the Yukawa coupling
\begin{equation}
 L_Y^o=(h_1 \bar{L_e}\phi^{(1)}+h_2 \bar{L_\mu}\phi^{(2)}+h_3 \bar{L_\tau}\phi^{(3)})N+\text{H.C.}+f N^T C N\Sigma\label{122}
\end{equation}
Although a term $M_N N^T CN$ is allowed in (\ref{122}) but it can be
absorbed in $\displaystyle{\frac{f\Lambda}{\sqrt{2}}}$ after the
symmetry breaking. After spontaneous symmetry breaking
$\langle\Sigma\rangle=\displaystyle{\frac{\Lambda}{2}}$(so that
$M=f\Lambda/\sqrt{2}$) and in the basis
$\bar{N}\left(\begin{tabular}{c}
               $\nu_e$\\
               $\nu_\mu$\\
               $\nu_\tau$
              \end{tabular}\right)
$, the Dirac neutrino mass matrix, $m_D^o$ is $3\times 1$ matrix
\begin{equation}
 m_D^o=\left(\begin{tabular}{c}
              $h_1$\\
              $h_2$\\
              $h_3$\\
             \end{tabular}\right)
\frac{v}{\sqrt{2}}
\end{equation}
The effective neutrino mass matrix is then
\begin{equation}
 M_\nu^o=m_D^o(m_D^o)^T\frac{1}{M}=\frac{v^2}{2 M}\left(
 \begin{tabular}{ccc}
  $h_1^2$ & $0$ & $0$\\
  $0$ & $h_2^2$& $0$\\
  $0$ & $0$ & $h_3^2$
 \end{tabular}\right)\label{135}
\end{equation}
Now $\Sigma^i$ carry flavor quantum numbers of $U_i$ but $\Sigma$
does not, being $[U(1)]^3$ singlet, and we may take
$\langle\Sigma^i\rangle\gg\langle\Sigma\rangle$ so that $M_1\gg M$
and then, as in clear from Eqs.(\ref{121}) and (\ref{135}), $\delta
M_\nu\ll M_\nu^0$. This also justifies neglect of mixing of $N$ with
$N^{(i)}$. In a way it is nice since in fermions mass hierarchy,
Yukawa couplings widely differ. Here it is due to $M_1\gg M$ while
Yukawa couplings are of same order. With $h_1\simeq h_2\simeq h_3$,
we have the required matrix given in Eq.(\ref{8}) or its variant to
be considered later, [c.f. Eq.(\ref{15})]. We may remark here that
although the Lagrangian (\ref{122}) as it stands is not flavour
blind, but if one takes $h_1=h_2=h_3=h$, then it is flavor singlet.
This is easy to see as follows. In three dimensional flavor space,
introduce two vectors $\textbf{L}=(L_1,L_2,L_3)$ and
$\boldsymbol{\Phi}=(\phi_1,\phi_2,\phi_3)$, then the Lagrangian
(\ref{122}) takes the form
\begin{equation}
L^0_Y=h[\bar{\textbf{L}}.\boldsymbol{\Phi}N+\text{H.C.}]+f N^T CN\Sigma\notag
\end{equation}
which is flavor single, just as
$(\boldsymbol{\bar{\Sigma}}.\boldsymbol{\pi})\Lambda$ is singlet
under isospin in hadron physics. When the symmetry is broken,
$M_\nu^0$ is then a multiple of unit matrix (flavor blind) and since
it dominates over $\delta M_\nu$, $M_\nu$ has approximate flavor
symmetry. It is in this limited sense that we talk about approximate
falvor symmetry.

We now explore the TBR possibility for the
diagonalization of the neutrino mass matrix given in Eq. (\ref{8}).
In spite of its attractiveness, its diagonalization is in conflict
with the neutrino data. It is instructive to show it as it would
provide us a guidance for possible modification of $M_{\nu}$ in Eq.
(\ref{8}) to obtain agreement with the experimental data. The
diagonalization of $M_{\nu}$ (we need to consider the diagonalization
of $\delta M_{\nu}$ as $m_{0}I$ commutes with any diagonalyzing
matrix) give among others the following relation \cite{9}
\begin{eqnarray}
\epsilon_{2}=\frac{\cos^{2}\theta_s-\tan^2\theta_r}{\sin^2\theta_s-\tan^2\theta_r}\epsilon_1,\quad\epsilon_1+\epsilon_2+\epsilon_3=0 \label{10}
\label{9}
\end{eqnarray}
where $\epsilon_i$ are the eigenvalues of $\delta M_\nu$.

Now
\begin{eqnarray}
m_{2}=m_{0}(1+\epsilon_2),\text{ } m_1=m_0(1+\epsilon_1),\text{
}m_3=m_0(1-\epsilon_1-\epsilon_2)\label{132}
\end{eqnarray}
The first of relations (\ref{132}) give
\begin{eqnarray}
\epsilon_+(1-2\tan^2\theta_r)=\epsilon_{-}\cos2\theta_s \label{11}
\end{eqnarray}
while
\begin{subequations}
\begin{eqnarray}
\Delta m^2_{12}=4m^2_0\epsilon_- \label{12a} \\
\Delta m^2=-6m^2_0\epsilon_+ \label{12b}
\end{eqnarray}
\end{subequations}
where
\begin{eqnarray}
\epsilon_+=\frac{\epsilon_2+\epsilon_1}{2},\text{
}\epsilon_-=\frac{\epsilon_2-\epsilon_1}{2} \label{14}
\end{eqnarray}

We require $\epsilon_-\ll\epsilon_+$ and the relation (\ref{11}) then
implies that $tan^2\theta_{r}\simeq 1/2$, contrary to the
experimental data. The relation (\ref{9}) is the consequence of
det$|\delta M_{\nu}|=0$. To avoid this, the simplest extension is
that
\begin{eqnarray}
\delta M_{\nu}=m_o\left(\begin{array}{ccc} a & \epsilon_{12} & \epsilon_{13} \\
\epsilon_{12} & 0 & \epsilon_{23} \\
\epsilon_{13} & \epsilon_{23} & 0
\end{array}\right) \label{15}
\end{eqnarray}
where $a\ll 1$. Then the relations in Eq. (\ref{9}) are replaced by
\begin{eqnarray}
a=-\epsilon_1
(\tan^2\theta_r-cos^2\theta_s)-\epsilon_2(\tan^2\theta_r-\sin^2\theta_s)\notag
\\
=\epsilon_+(1-2\tan^2\theta_r)-\epsilon_-\cos2\theta_s \label{16}
\end{eqnarray}
\begin{eqnarray}
\epsilon_{1}+\epsilon_{2}+\epsilon_{3}=a \label{17}\end{eqnarray}
Further while the relation (\ref{12a}) remains the same but
(\ref{12b}) is changed to
\begin{subequations}
\begin{eqnarray}\
\Delta m^2&=&m_0^2[2a-6\epsilon_+]\notag \\
&=&-4m_0^2[\epsilon_+(1+\frac{1}{2}\tan^2\theta_r)+\frac{1}{2}\epsilon_-\cos2\theta_s)]\label{18a}
\end{eqnarray}
On using Eq. (\ref{12a})
\begin{eqnarray}
|\Delta
m^2|=4m_0^2\epsilon_+(1+\tan^2\theta_r)+\frac{1}{2}\cos\theta_s\Delta
m_{12}^2\notag
\end{eqnarray}
Since $|\Delta m^2|\gg\Delta m_{12}^2$, it follows that
\begin{eqnarray}
|\Delta m^2|\simeq4m_0^2|\epsilon_{+}| \label{18b}
\end{eqnarray}
\end{subequations}
Thus from Eqs. (\ref{12a}), (\ref{18b}) and (\ref{4})
\begin{eqnarray}
\left|\frac{\epsilon_-}{\epsilon_+}\right|=\frac{\Delta
m_{12}^2}{|\Delta m^2|}=3.2\times 10^{-2} \label{19}
\end{eqnarray}

The other relations which diagonalization give are
\begin{eqnarray}
\cos2\theta_a\epsilon_+-\sin2\theta_a\sin2\theta_s\sin\theta_r\epsilon_-=0
\label{20}
\end{eqnarray}
\begin{eqnarray}
\epsilon_{12}+\epsilon_{13}=\cos\theta_r\sin2\theta_s(\cos\theta_a-\sin\theta_a)\epsilon_--2(\cos\theta_a+\sin\theta_a)\tan\theta_r\epsilon_+\label{21}
\end{eqnarray}

\begin{eqnarray}
\epsilon_{12}-\epsilon_{13}=\cos\theta_r\sin2\theta_s(\cos\theta_a+\sin\theta_a)\epsilon_-+2(\cos\theta_a-\sin\theta_a)\tan\theta_r\epsilon_+
\label{22}
\end{eqnarray}
\begin{eqnarray}
\epsilon_{23}=-\sin2\theta_a(\epsilon_++\epsilon_-\cos2\theta_s)-\epsilon_-\cos2\theta_a\sin2\theta_s\sin\theta_{r}
\label{23}
\end{eqnarray}

To proceed further it is convenient to use the expansion about the
maximum atmospheric angle $\sin^2\theta_a=\frac{1}{2}$
\begin{eqnarray}
\sin\theta_a=-\frac{1}{\sqrt{2}}(1+t),\text{
}\cos\theta_a=\frac{1}{\sqrt{2}}(1-t),\text{ }\sin^2\theta_a=0.5+t
\label{24}
\end{eqnarray}
Then the relations (\ref{20}-\ref{23}) simplify to
\begin{eqnarray}
-2t\epsilon_++\sin2\theta_s\sin\theta_r\epsilon_-&=&0 \label{25} \\
\epsilon_{12}-\epsilon_{13}&=&-\sqrt{2}t\cos\theta_r\sin2\theta_s\epsilon_-+2\sqrt{2}\tan\theta_r\epsilon_+
\label{26}
\\
\epsilon_{23}&=&\epsilon_++\epsilon_-\cos2\theta_s \label{27} \\
\epsilon_{12}+\epsilon_{13}&=&\sqrt{2}\cos\theta_r\sin2\theta_s\epsilon_-+2\sqrt{2}t\tan\theta_r\epsilon_+
\notag \\
&=&\sqrt{2}\sin2\theta_s(1-\frac{3}{2}\sin^2\theta_r)\epsilon_-
\label{28}
\end{eqnarray}
where in the second step we have used Eq. (\ref{25}). Further from
Eqs. (\ref{16}) and (\ref{27})
\begin{eqnarray}
a-\epsilon_{23}=-2\epsilon_+\tan^2\theta_r-2\epsilon_-\cos2\theta_s
\label{29}
\end{eqnarray}
so that together with Eq. (\ref{28})
\begin{eqnarray}
\tan2\theta_s=\frac{-\sqrt{2}(\epsilon_{12}+\epsilon_{13})(1+\frac{3}{2}\sin^2\theta_r)}{a-\epsilon_{23}+2\tan^2\theta_r\epsilon_+}
\label{30}
\end{eqnarray}

We note that if
$\nu_{\mu}\leftrightarrow\nu_{\tau}(2\leftrightarrow3)$ symmetry is
imposed so that $\epsilon_{12}=\epsilon_{13}$, $t\rightarrow0$,
$\theta_r\rightarrow0$, the relations (\ref{25}), (\ref{26}) are
identically satisfied. Further if $a-\epsilon_{23}=-\epsilon_{12}$,
then Eq. (\ref{30}) gives $\tan2\theta_s=2\sqrt{2}$ i.e. the TB
solar angle.

However if $\sin\theta_r\neq0$, Eq. (\ref{25}) implies that
\begin{eqnarray}
\sin\theta_r=\frac{2t}{\sin2\theta_s}\frac{\epsilon_+}{\epsilon_-}\notag \\
\simeq\pm \frac{1}{\sqrt{2}}t\times10^{2}, \label{31}
\end{eqnarray}
on using Eq. (\ref{19}) and the TB value of $\theta_s$. This can
accommodate any finite value of $\sin\theta_r$ for extremely small
deviation from the maximal atmospheric angle
$\sin^2\theta_a=\frac{1}{2}$.
\begin{figure}
 \includegraphics{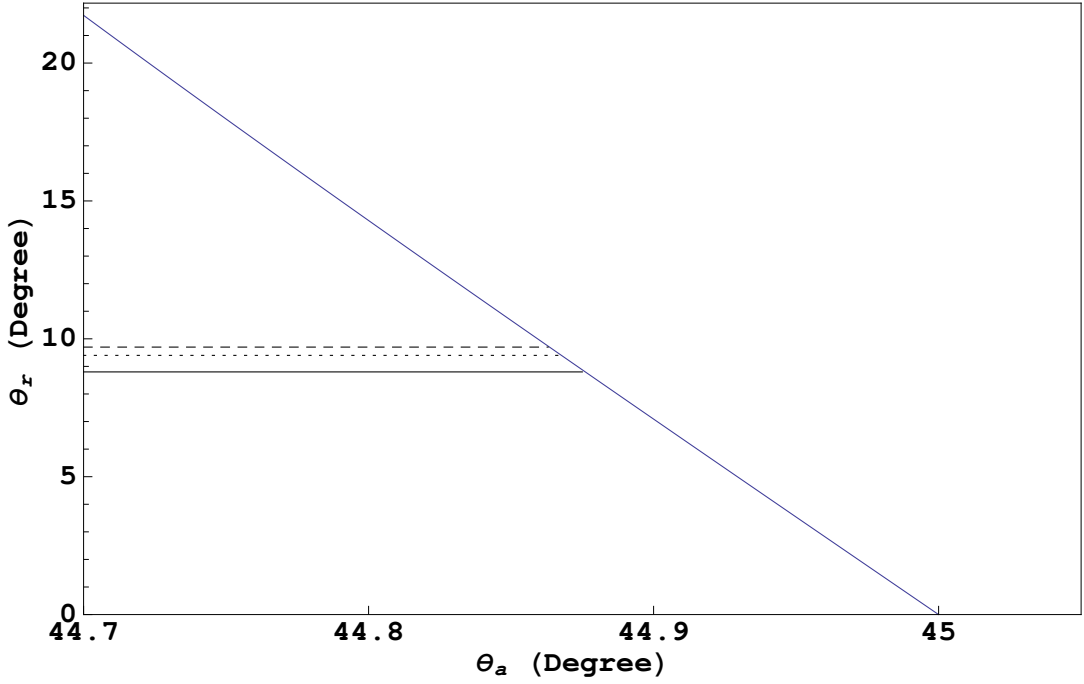}
\caption{$\theta_r$ as a function of $\theta_a$. The solid, dotted and dashed lines respectively correspond to the best fit values for DAYA BAY, RENO and T2K collaborations.}
\end{figure}
Putting $t$ as given in the third relation of Eq.(\ref{24}) in
Eq.(\ref{31}), we plot $\theta_r$ as a function of $\theta_a$ in
Fig.1 which is if sign is taken negative in Eq.(\ref{31}). If the
sign is positive, $\theta_{a}$ is greater than $45^{\circ}$ and for
$\theta_{r}=10^{\circ}$ it is $45.1^{\circ}$. It is clear that one
can achieve $\theta_r\simeq 7^\circ$ to $ 10^\circ$, keeping
$\theta_a$ around $45^\circ$. Thus it covers recently measured
values of $\theta_r$ by $T2K$, DAYA BAY\cite{5} and RENO
collaboration\cite{6}. It follows from Eq. (\ref{30}) that TB solar
angle is obtained i.e. $\tan2\theta_s=2\sqrt{2}$, if
\begin{eqnarray}
a-\epsilon_{23}+2\tan^2\theta_r\epsilon_+=-\frac{1}{2}(\epsilon_{12}+\epsilon_{13})(1+\frac{3}{2}\sin^2\theta_r)
\label{32}
\end{eqnarray}

It remains to determine the parameters in the $\delta M_{\nu}$ given in Eq. (\ref{15}). It follows from Eqs. (\ref{18b}), (\ref{27}), (\ref{28})
and (\ref{29}) that
\begin{eqnarray}
|\epsilon_{23}|\simeq a\simeq|\epsilon_+|=\frac{1}{4}\frac{|\Delta
m^2|}{m_0^2} \label{33} \\
\frac{\epsilon_{12}+\epsilon_{13}}{\epsilon_+}\simeq
\sqrt{2}\sin2\theta_s\frac{\epsilon_-}{\epsilon_+}\simeq4\times10^{-2}
\label{34}
\end{eqnarray}

On the other hand from Eqs. (\ref{5}) and (\ref{26})
\begin{eqnarray}
\left|\frac{\epsilon_{12}-\epsilon_{13}}{\epsilon_+}\right|\simeq
2\sqrt{2}\tan\theta_r\simeq4\times10^{-1} \label{35}
\end{eqnarray}

Thus it is possible to have TB solar angle
$\sin^2\theta_s=\frac{1}{3}$ and almost maximal atmospheric angle
$\sin^2\theta_a\simeq\frac{1}{2}$ and non zero $\sin^2\theta_r$ but
at the cost of $\nu_{\mu}\rightarrow\nu_{\tau}$ symmetry as the
relations (\ref{34}) and (\ref{35}) would imply. Finally the
oscillation data give only a lower bound on the heaviest of the
neutrino mass $m_h\geq|\Delta m^2|>0.05$eV but cannot fix it.
However $m_0$ is further constrained by WMAP data, $\sum
m_{i}<(0.4-0.7)$eV. Taking $m_{0}\simeq0.1$eV, we get from Eqs.
(\ref{4}), (\ref{33}), (\ref{34}) and (\ref{35}) that
\begin{eqnarray}
|\epsilon_{23}|\simeq a\simeq |\epsilon_+|\simeq6\times10^{-2}\notag \\
\epsilon_{12}+\epsilon_{13}\simeq2.4\times10^{-3}\notag \\
|\epsilon_{12}-\epsilon_{13}|\simeq2.4\times10^{-2}\notag
\end{eqnarray}
Then
\begin{eqnarray}
\epsilon_{12}\simeq-\epsilon_{13}\simeq10^{-2} \label{36}
\end{eqnarray}
if the sign is chosen to be positive for
$\frac{\epsilon_-}{\epsilon_+}$. If the sign is negative,
$\epsilon_{12}$ and $\epsilon_{13}$ should be interchanged. This is
meant for the rest of the manuscript.

 We may remark here that there are four parameters, apart from $m_o$, in
Eq.(\ref{15}). There are two mass differences and three mixing
angles. Thus the prediction one gets is a relationship between
$\theta_r$ and $\theta_a$ which is shown in Fig.1. All above
parameters as well Fig.1 are obtained by using best fit values given
in Eq.(\ref{4}).

All the values given in Eq.(\ref{36})are consistent with small perturbations (at
least an order of magnitude smaller) to $M_{\nu}=m_0I$, $I$ being
the unit matrix. It is important to note that Eq. (\ref{36}) implies
approximate $\nu_{\mu}\rightarrow -\nu_{\tau}$ symmetry (instead of
$\nu_{\mu}\rightarrow\nu_{\tau}$ symmetry,which would imply
$\epsilon_{12}-\epsilon_{13}=0$). On the other hand the exact
$\nu_\mu\rightarrow -\nu_\tau$ symmetry would imply
$\epsilon_{12}=-\epsilon_{13}$ i.e. $\epsilon_{12}+\epsilon_{13}=0$.
In order to see how this happens for $\theta_{r}\neq 0$, we note from
Eqs. (\ref{19}), (\ref{34}) and (\ref{35}) that
\begin{equation}
\frac{\epsilon_{13}}{\epsilon_{12}}=\frac{1-\frac{1}{\sqrt{2}}10^2
\tan(\theta_r)}{1+\frac{1}{\sqrt{2}}10^2 \tan(\theta_r)}
\end{equation}

\begin{figure}[ht]
 \includegraphics{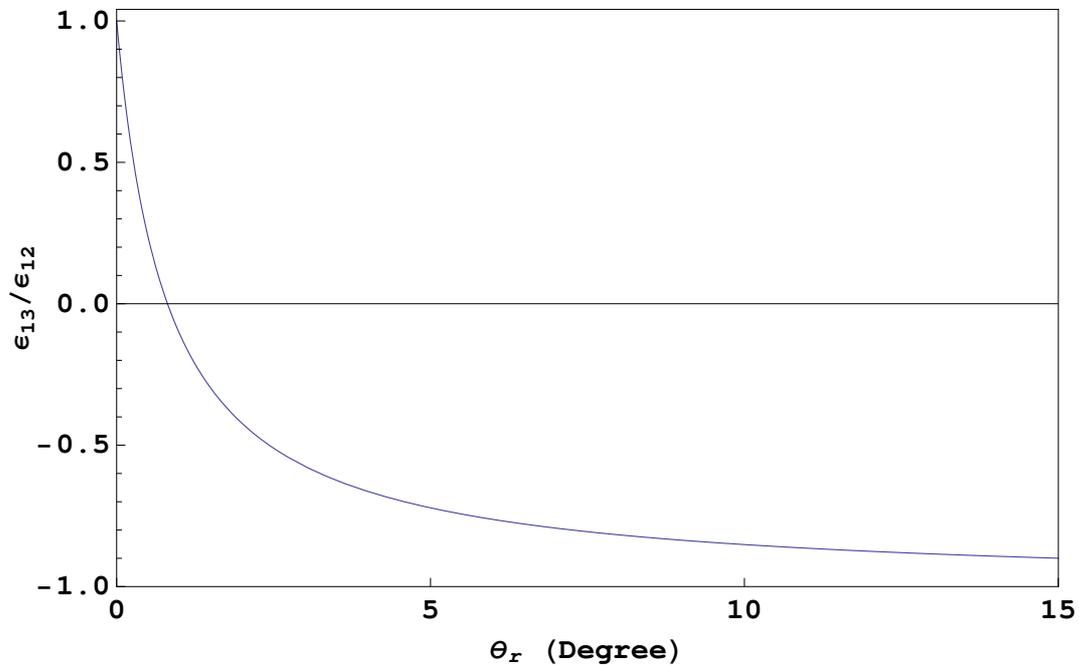}
\caption{$\displaystyle{\epsilon_{13}/\epsilon_{12}}$ as a function
of $\theta_r$. }
\end{figure}
In Fig. 2, we plot $\epsilon_{13}/\epsilon_{12}$ as a function of
$\theta_r$. We see how the transition from exact $\nu_\mu\rightarrow
\nu_\tau$ symmetry $(\epsilon_{12}/\epsilon_{13}=1)$ at $\theta_r=0$
to approximate $\nu_{\mu}\rightarrow-\nu_{\tau}$ symmetry
$\left(\displaystyle{\frac{\epsilon_{13}}{\epsilon_{12}}}\simeq -1,
\text{e.g. at } \theta_r\simeq
10^\circ,\displaystyle{\frac{\epsilon_{13}}{\epsilon_{12}}}\approx
-0.85\right)$ takes place.

\section{Summary and Conclusion}

We have considered a model of approximate flavor symmetry which has
near degenerate neutrino mass. It is shown that it is possible to
have nonzero reactor angle while preserving the TB solar angle
$\sin^2\theta_s=\frac{1}{3}$ and near maximal atmospheric angle
$\sin^2\theta_a\simeq0.5$.The non-zero $\sin\theta_r$ implies
approximate $\nu_{\mu}\rightarrow-\nu_{\tau}$ symmetry rather than
that $\nu_{\mu}\rightarrow\nu_{\tau}$ symmetry.

\section*{Acknowledgments}
The author would like to thank Prof. Fernando Quevedo for
hospitality at the Abdus Salam International Centre for Theoretical
Physics (AS-ICTP), Trieste, where the work was performed.


\begin{thebibliography}{111}

\bibitem{1} B. Pontecorvo, Sov. Phys. JETP \textbf{6} (1957) 429 [Zh. Eksp.
Teor. Fiz. 33 (1957) 549]; Z. Maki, M. Nakagawa and S. Sakata, Prog.
theor. Phys. \textbf{28} (1962) 870.

\bibitem{2} K. Nakamura et al. (Particle Data Group), J. Phys. \textbf{G 37} (2010) 075021.

\bibitem{3} T. Schwertz, M. Tortala, and J. W. F Valle, [arXiv: 1108.1376]; G. L. Fogli, E. Lisi, A. Marrone, A. Palazzo and A. M. Rotunno, [arXiv:1106.6028].

\bibitem{4} K. Abe et al. (T2K collaboration), arXiv:1106.2822.

\bibitem{5} F. P. An \textit{et al.} (Daya Bay Collobation), arXiv: 1203.1669[hep-ex].

\bibitem{6} J.K. Ahn et al (RENO colleboration): arXiv: 1204.0626[hep-ex].

\bibitem{7} Riazuddin, Phys. Rev. \textbf{D 81} (2010) 057301.

\bibitem{8} For a review see, for example, R. N.Mohapatra and A.Y. Smirnov: arXiv:0603118, Ann Rev. Nucl. Part. Sci 59 (2006) 569 .

\bibitem{9} C.S. Lam, Phys. Lett. \textbf{B 507} (2001) 214 ; Phys. Rev. \textbf{D 71} (2003) 093001

\bibitem{10} P.F. Harrison, D. H. Perkins and W. G. Scott, Phys. Lett. \textbf{B 530} (2002) 167

\bibitem{11} S. F. King, arXiv:0903.3199v5.

\bibitem{12} T. Araki, Phys. Rev. \textbf{D 84} (2011) 037301 , arXiv: 1106.5211

\bibitem{13} S. Morisi, K. M. Patel and E. Peinado, Phys. Rev. \textbf{D 84} (2011) 053002, arXiv: 1107.0696.

\bibitem{14} S. F. King and C. Luhn, arXiv: 1112.1959.

\bibitem{15} S. L. Glashow, arXiv: 0912.4976.

\bibitem{16} Fayyazuddin and Riazuddin, Phys. Rev. \textbf{D 35} (1987) 2201; Riazuddin, JHEP \textbf{10} (2003) 009; Riazuddin, Phys. Rev. \textbf{D 63} (2001) 033003.

\bibitem{17} A. Zee, Phys. Lett. \textbf{B 93} (1980) 389. For a recent discussion of this model, see ref. [16].

\bibitem{18} Xiao-Gang He and S.K.Majee, arXiv: 1111.2293v.2

\bibitem{19} J.Matsuda, C.J.Jarlskog, S.Skadhauf and M.Tanimoto, hep-ph0005147; Riazuddin, 3rd reference in [13].

\end{thebibliography}
\end{document}